\documentclass[twocolumn,10pt]{asme2e}

\usepackage{xcolor}
\usepackage{mathtools}
\usepackage{gensymb}
\usepackage{arydshln}
\usepackage{colortbl}
\confshortname{DSCC 2020}
\conffullname{the ASME 2020 \\ Dynamic Systems and Control Conference}

\confdate{4-7}
\confmonth{October}
\confyear{2020}
\confcity{Pittsburgh, PA}
\confcountry{USA}

\papernum{DSCC2020-3283}

\title{An improved model of height profile for Drop-on-demand print\\ of UV curable ink}

\author{Yumeng Wu
    \affiliation{
	School of Mechanical Engineering\\
	Purdue University\\
	West Lafayette, Indiana 47907\\
    Email: wu350@purdue.edu
    }	
}

\author{George Chiu\thanks{Address all correspondence to this author.} 
    \affiliation{
	School of Mechanical Engineering\\
	Purdue University\\
	West Lafayette, Indiana 47907\\
    Email: gchiu@purdue.edu
    }
}

\begin{document}
\maketitle
\begin{abstract}
This paper proposes an improved  model of height profile for drop-on-demand printing of UV curable ink.
Unlike previous model, the proposed model propagates volume and covered area based on height difference between adjacent drops.
Height profile is then calculated from the propagated volume and area. 
Measurements of 2-drop and 3-drop patterns are used to 
experimentally compute model parameters.
The parameters are used to predict and validate height profiles of 4 and more drops in a straight line.
Using the same root mean square (RMS) error as benchmark, 
this model achieves 5.9\% RMS height profile error on 4-drop lines.
This represents more than 60\% reduction from graph-based model and an improvement from our previous effort.
\end{abstract}

\begin{nomenclature}
\entry{$a_{i j}[k]$}{Area covered by ink at cell $(i,j)$ after the $k^{th}$ drop.}
\entry{$v_{i j}[k]$}{Volume at cell $(i,j)$ after the $k^{th}$ drop.}
\entry{$h_{i j}[k]$}{Average height at cell $(i,j)$ after the $k^{th}$ drop.}
\entry{$A[k]$}{Area matrix in the area of interest after the $k^{th}$ drop.}
\entry{$V[k]$}{Volume matrix in the area of interest after the $k^{th}$ drop.}
\entry{$H[k]$}{Height matrix in the area of interest after the $k^{th}$ drop.}
\entry{$\Delta a_{i j}[k]$}{Area change at cell $(i, j)$ due to the $k^{th}$ drop.}
\entry{$\Delta v_{i j}[k]$}{Volume change at cell $(i, j)$ due to the $k^{th}$ drop.}
\entry{$\Delta A[k]$}{Matrix of area change due to the $k^{th}$ drop.}
\entry{$\Delta V[k]$}{Matrix of volume change due to the $k^{th}$ drop.}
\entry{$M_a[k]$}{Impact matrix of height on the $k^{th}$ drop area distribution.}
\entry{$M_v[k]$}{Impact matrix of height on the $k^{th}$ drop volume distribution.}
\entry{$M_a$}{Matrix characterizing impact of height on area distribution.}
\entry{$M_v$}{Matrix characterizing impact of height on volume distribution.}
\entry{$S$}{One step shift matrix.}
\entry{$c$}{The ratio between single drop volume and single cell area.}
\entry{$d$}{Nominal pitch between two drops.}
\entry{$r$}{Radius of single drop's footprint on substrate.}
\entry{$v_s$}{Total volume of a single drop.}
\end{nomenclature}

\section{\MakeUppercase{Introduction}}

Additive manufacturing (AM) or 3-D printing,
has expanded from prototyping into manufacturing \cite{wohlers2013wohler}.
With additive manufacturing,
small to median batch production is now economically viable. 
This is due to its low material waste and cost, in comparison to the traditional computer numerical control (CNC) machining \cite{ford2016additive}.
Many industries, such as aerospace,
automotive and biomedical will be affected by the application of additive manufacturing \cite{kupper2017get}.

Among different additive manufacturing processes,
drop-on-demand printing is more common with biomedical applications,
since it can fulfill  the demands on both geometry and functions \cite{simpson2016preparing}.
Moreover,
drop-on-demand printing accepts wide range of inks and can achieve high resolution \cite{cooley2002applicatons,sirringhaus2003inkjet}.
The technology advancement and cost reduction from additive manufacturing make customized  medical procedures and devices available.
An endovascular aneurysm repair (EVAR) can be customized to meet patient-specific needs with additive manufacturing  \cite{lei2020new}.
Hearing aids are manufactured with 3-D printing to better fit the patients' ear profiles \cite{chu2008design}. 

Many researchers have developed functional materials specifically  for drop-on-demand 3-D printing \cite{gantenbein2018three}.
However, there is limited work focusing on meeting the geometry demand.
A single drop is usually modeled as a spherical cap after solidification \cite{Doumanidis2000}.
Starting from the second drop, 
drop profile is no longer easy to obtain,
because the substrate profile with drops is not flat anymore. 
Some researchers use computational methods to model the profile \cite{Xu_Basaran_2007,choi2017numerical,gunjal2003experimental}.
Although these models are accurate, 
the high computational cost  makes them less desirable to process in real time and to be used for process control.
Other researchers simplify the model to balance between accuracy and computation demand.
The graph-based model is proposed to capture the dynamics of height propagation, 
but the error is often greater than 10\% \cite{Guo2018}.
A model based on volume and area propagation reduces error in height profile, 
but it maybe oversimplified \cite{WuY.2019Mhpf}. 

In this paper,
the previous model proposed in \cite{WuY.2019Mhpf} is extended to quantify the impact of height difference on material flow.
The model calculates height profile indirectly from volume and area propagation.
Both volume and area propagations include the impact due to height difference prior to deposition.
The remaining of the paper is organized as follow.
the height profile is introduced in Section \ref{sec:single}.
Propagation model is introduced in Section \ref{sec:model}.
Experimental validation is included in Section \ref{sec:exp_valid}.
Lastly, Section \ref{sec:conclusion} is the conclusion.

\section{\MakeUppercase{Single Drop Height Profile Model}}
\label{sec:single}

\begin{figure}
	\centering
	\includegraphics[width=\linewidth]{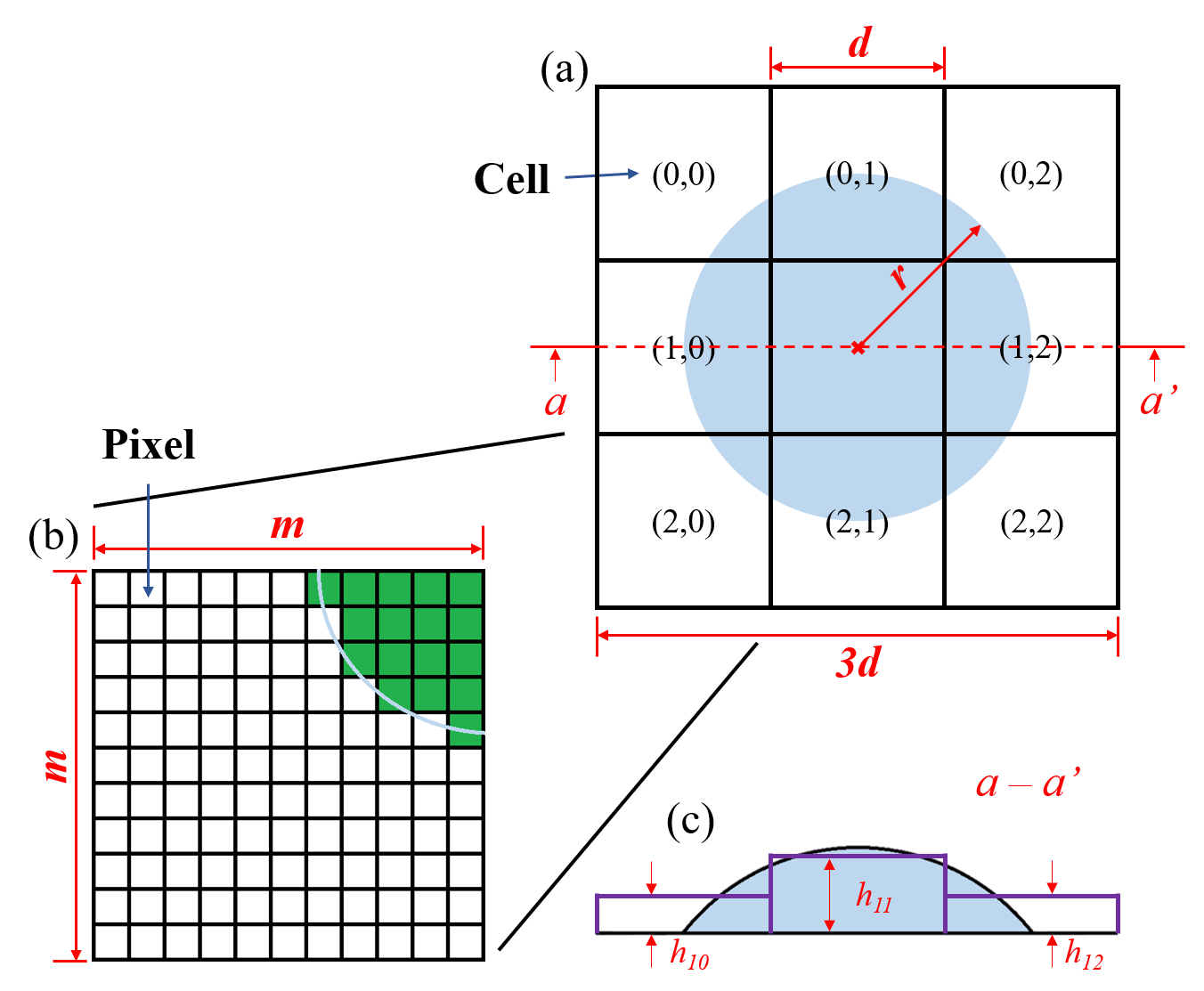}
	\caption{(a): Single drop with radius $r$ on $3 \times 3$ cell. \hspace{\textwidth} 
	(b): Zoomed-in section of cell (2,0) in (a), illustrating how to count area covered by inks.
	(c): Height profile comparison of center row. Proposed height profile is in purple, while spherical cap height profile is in light blue.}
	\label{fig:single}
\end{figure}

\mbox{Figure \ref{fig:single}} illustrates a single drop height profile model. 
\mbox{Figure \ref{fig:single}.(a)} shows $3 \times 3$ square cells,
where a drop of material is deposited at the center of the $3 \times 3$ cells. 
The size of each cell is $d \times d$, where $d$ is the pitch distance  between two adjacent drops. 
The $k^{th}$ drop is deposited at the center of cell $(1,k)$.
Such $3 \times 3$ cells are denoted as $B_k$,
which is the region of interest of the $k^{th}$ drop and can be defined as
\begin{equation}
\label{eq:bk}
   B_k =  \{(i,j)|i \in \{0,1,2\}, j \in \{k-1,k,k+1 \} \}.
\end{equation}

Assuming a single drop deposited on a flat non-porous surface  forms a spherical cap with radius $r$, if $0.5 d \le r <1.5 d$,
the drop covers the $3 \times 3$ cells,
see the shaded area in \mbox{Fig. \ref{fig:single}.(a)}. 
This range of pitch distance is used to make enough overlap between adjacent drops to ensure print quality. 

Each cell in $B_k$ is indexed by $(i,j)$,
where $i$ is the row index and $j$ is the column index.
Both $i$ and $j$ start from 0.

$H[k]$ denotes the height matrix representing the cell height within region of interest after the $k^{th}$ drop is deposited at cell $(1,k)$. 
For the first drop,
the region of interest is $\left\{(i,j) | i,j \in \{0,1,2\}\right\}$.
After the first drop is deposited on cell  (1,1),
the height matrix can be written as
\begin{equation}
\label{eq:h_k}
    H[1] = \left[ \begin{array}{ccc}
       h_{0 0}[1] &  h_{0 1}[1] &  h_{0 2}[1]\\
       h_{1 0}[1] &  h_{1 1}[1] &  h_{1 2}[1]\\
       h_{2 0}[1] &  h_{2 1}[1] &  h_{2 2}[1]  
\end{array} \right],
\end{equation}
where $h_{ij}[1]$ denotes the height at cell $(i,j)$ in $B_1$ after the first drop is deposited at cell $(1,1)$,
which is computed from the ratio between the drop volume within the cell and the area covered by the drop within the cell.
Figure \ref{fig:single}.(c) illustrates the height profiles of cells (1,0), (1,1) and (1,2).
The shaded portion in \mbox{Fig. \ref{fig:single}.(a)} 
represents the height profile from the spherical cap model.
Horizontal solid lines represent the proposed height profile.

A Zeta-20 optical profilometer is used to obtain the height profile experimentally. 
Its spatial resolution makes measuring height profile possible. 
The measurements from the profilometer of cell (2,0) in \mbox{Fig. \ref{fig:single}.(a)} is a high resolution image with $m \times m$ pixels of height measurements.
This is illustrated in \mbox{Fig. \ref{fig:single}.(b)}.
The area of each pixel $(a_c)$ is $d^2/m^2$. 
Pixels covered by ink are shaded. 
The cell area $(a_{ij}[k])$ is the ratio between area covered by the ink and total cell area,
If there are $M$ pixels covered by ink,
the cell area $(a_{ij}[k])$ is 
\begin{equation}
    \label{eq:cell_area_pct}
    a_{ij}[k] = \frac{M}{m^2} \times 100\%.
\end{equation}
Similar to $H[1]$,
the aggregated area matrix within region of interest after the first drop, $B_1$,  can be written as
\begin{equation}
\centering
\label{eq:a_single}
    A[1] = \left[ \begin{array}{ccc}
       a_{0 0}[1] &  a_{0 1}[1] &  a_{0 2}[1]\\
       a_{1 0}[1] &  a_{1 1}[1] &  a_{1 2}[1]\\
       a_{2 0}[1] &  a_{2 1}[1] &  a_{2 2}[1]  
\end{array} \right],
\end{equation}
where $a_{i j}[1]$ denotes the percentage area covered by the first drop in cell $(i, j)$.
With $i_m,j_m=0,\cdots,m-1$ representing pixel indices within a cell, 
the pixel height $(h_{i_m j_m})$ is zero if the pixel is not covered by ink.
Otherwise, the pixel height is non-zero.

The ink volume at each pixel $(v_{i_m, j_m})$ is the product of pixel area and height,
$v_{i_m, j_m} = a_c h_{i_m,j_m}$.
The  cell volume at $(i,j)$ after the $k^{th}$ drop, $v_{ij}[k]$ is the percentage of single drop volume.
The single drop volume, $v_s$, is a constant. 
For each cell $(i,j)$,
after the first drop is deposited at (1,1), the volume $(v_{ij}[1])$ can be computed by
\begin{equation}
    \centering
    \label{eq:single}
    \begin{split}
        v_{i j}[1] &=  \frac{\sum_{i_m=0}^{m-1}\sum_{j_m=0}^{m-1} v_{i_m, j_m}}{v_s}  \times 100\%\\
        &= \frac{\sum_{i_m=0}^{m-1}\sum_{j_m=0}^{m-1} h_{i_m j_m} a_c}{v_s} \times 100\%. 
    \end{split}
\end{equation}

\begin{figure}
    \centering
    \includegraphics[width=0.95\linewidth]{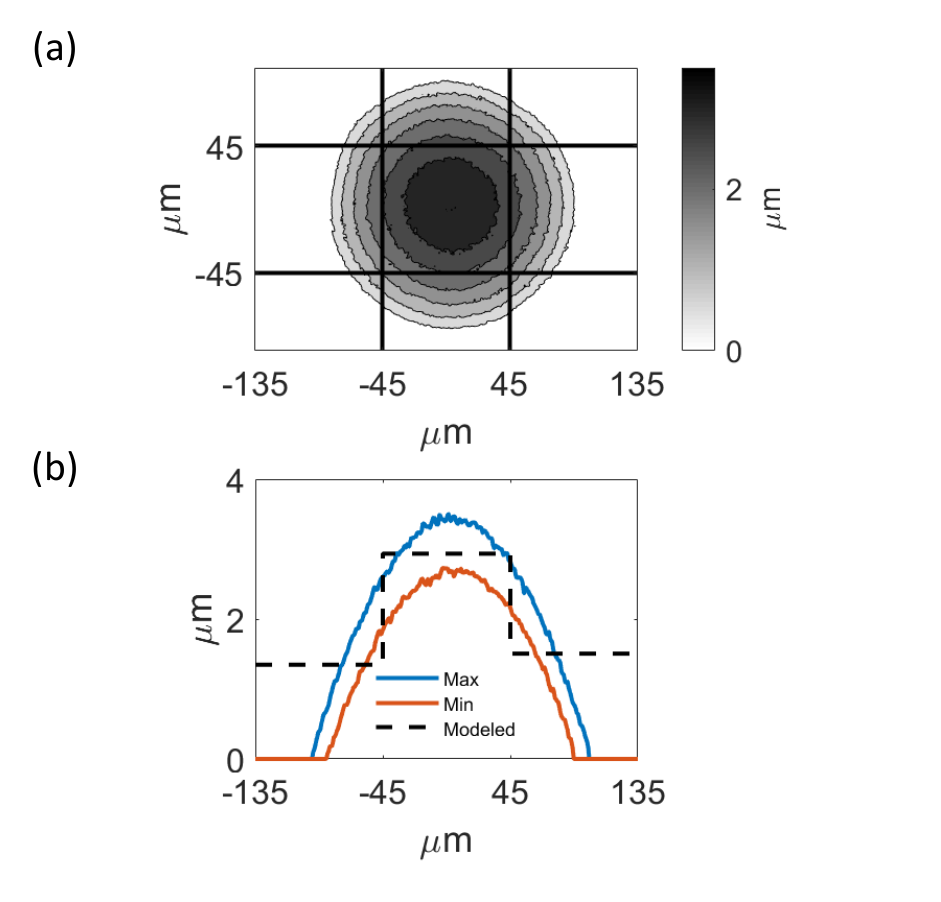}
    \caption{(a): The contour of single drop, marked with 90 $\mu m$ pitch size;
    (b): Solid lines are the measurements of the sample shown in (a) along the center row.
    Blue solid line is the maximum height along the center row;
    red solid line is the minimum height along the center row;
    black dashed line is the cell height of all samples along the center row. }
    \label{fig:s_contour}
\end{figure}

Similarly, the aggregated volume matrix within $B_1$ can be written as
\begin{equation}
\centering
\label{eq:v_single}
    V[1] = \left[ \begin{array}{ccc}
       v_{0 0}[1] &  v_{0 1}[1] &  v_{0 2}[1]\\
       v_{1 0}[1] &  v_{1 1}[1] &  v_{1 2}[1]\\
       v_{2 0}[1] &  v_{2 1}[1] &  v_{2 2}[1]  
\end{array} \right].
\end{equation}
With $a_{i j}[1]$ and $v_{i j}[1]$ both obtained, $h_{i j}[1]$ can be calculated as
\begin{equation}
    \centering
    \label{eq:h_single}
    \begin{split}
    h_{i j}[1] &= c \frac{v_{i j}[1]}{a_{i j}[1]} = c \frac{\sum_{i_m=0}^{m-1}\sum_{j_m=0}^{m-1} h_{i_m j_m}}{M}, \\ \text{and \hspace{0.1in} }
        c & = \frac{v_s}{m^2 a_c},
    \end{split}
\end{equation}
where $c$ is a constant,
calculated from the ratio between the single drop volume and single cell area. 

\subsection{EQUIPMENT AND MEASUREMENTS}
The single drop height profile is obtained from experiments.
Besides a Zeta-20 optical profilometer, 
the experimental system also includes a PI Precision XY Stage,
a Microdrop piezo-electrical dispensing system with heated nozzle and a UV light.
With 50x objective lens and 0.35x coupler in the profilometer, the Z-resolution is 0.04 $\mu m$ and the pixel area $(a_c)$ is 0.49 $\mu m^2$.
The constant $c$ for this setup is 7.0751.
The dispensing nozzle is mounted vertically on a mechanical linear stage.
The UV inks (C-Flex Cyan) are made by Kao Collins Inc.
The system is programmed to cure the drop immediately after deposition on a microscope slide for 1 second under the UV light,
which is mounted near the dispensing nozzle. 

Each height profile is derived by averaging volume and area from multiple samples,
i.e. the height profile is not identical to the average of height profiles from all samples.
Cell area $a_{i j}[1]$ and volume $v_{i j}[1]$ of each sample are obtained experimentally,
before calculating $h_{i j}[1]$ from \mbox{Eq. (\ref{eq:h_single})}.
$a_{ij}[1]$ is the percentage area covered within cell $(i,j)$.
$v_{i j}[1]$ is the percentage volume with respect to the single-drop volume.
Figure \ref{fig:s_contour}.(a) shows the contour of one sample with black grid representing the cell boundaries. 
The blue and red solid lines in \mbox{Fig. \ref{fig:s_contour}.(b)} show the maximum and minimum height along the center row of the sample in (a),
respectively, 
and the black dashed line shows the cell height $h_{1j}[1]$ from all single-drop samples.

\begin{figure*}[ht] 
  \begin{minipage}[b]{0.5\linewidth}
    \centering
    \includegraphics[width=0.85\linewidth]{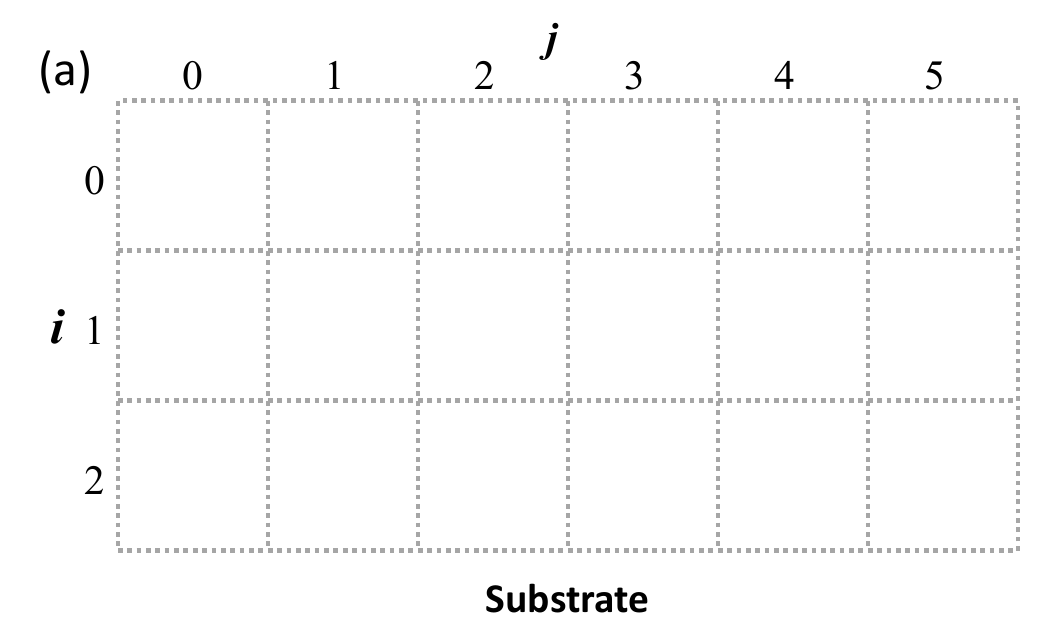}
    \vspace{4ex}
  \end{minipage}
  \begin{minipage}[b]{0.5\linewidth}
    \centering
    \includegraphics[width=0.85\linewidth]{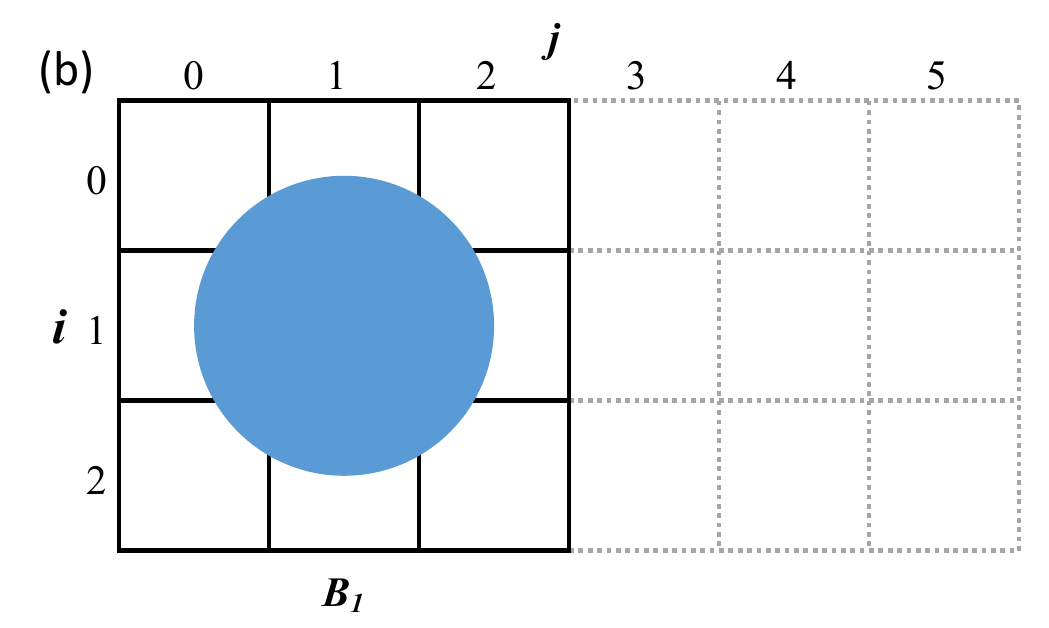}
    \vspace{4ex}
  \end{minipage} 
  \begin{minipage}[b]{0.5\linewidth}
    \centering
    \includegraphics[width=0.85\linewidth]{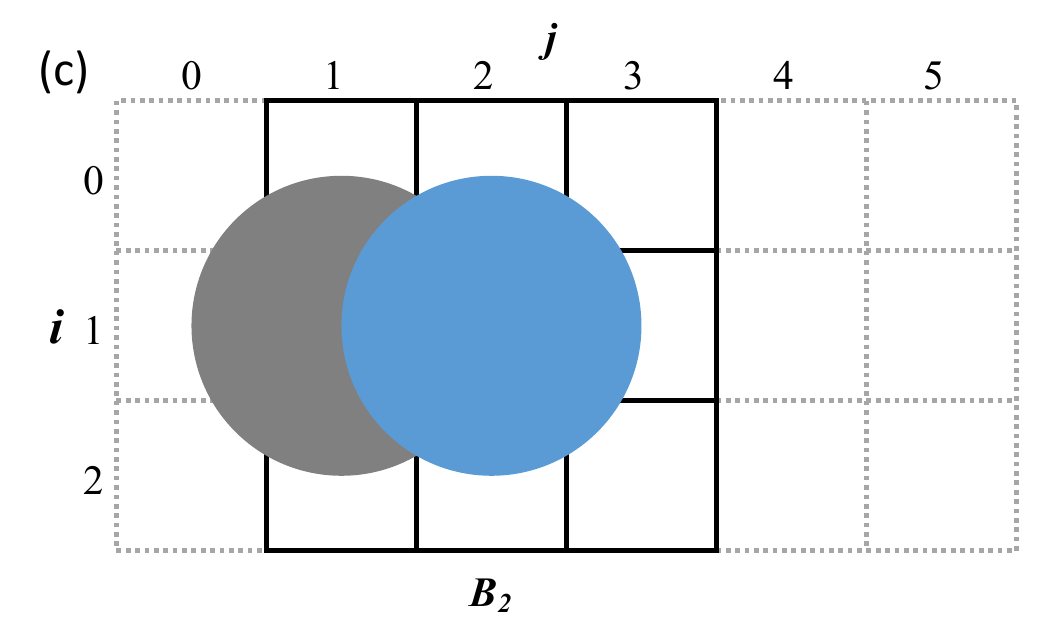}
    \vspace{4ex}
  \end{minipage}
  \begin{minipage}[b]{0.5\linewidth}
    \centering
    \includegraphics[width=0.85\linewidth]{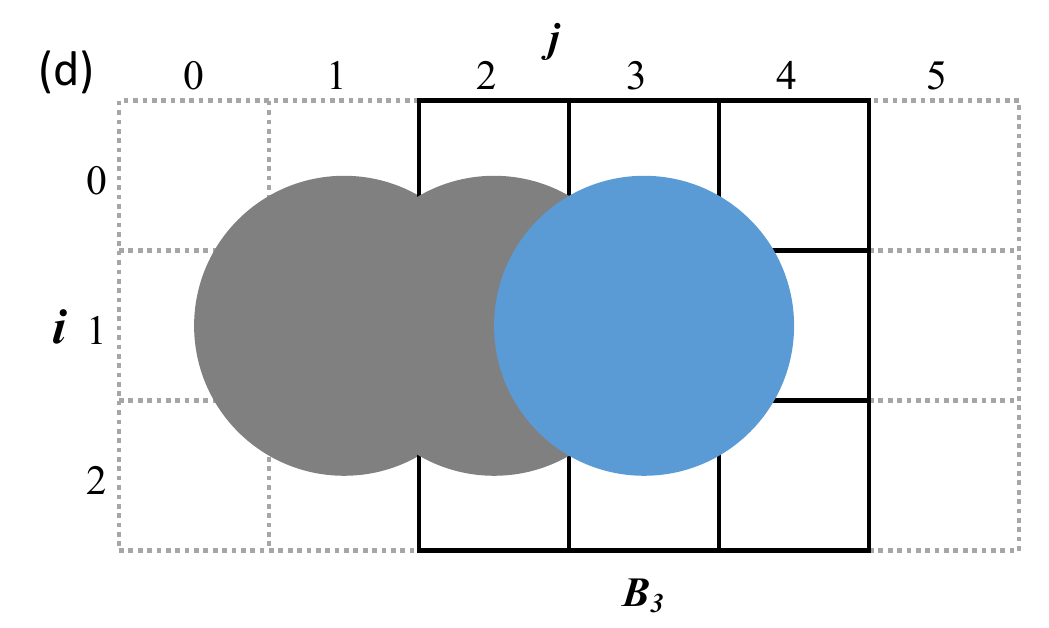}
    \vspace{4ex}
  \end{minipage}
  \caption{From upper-left to lower-right: (a) illustrates the cell-grid of 4-drop straight line before printing, or substrate;
  (b) illustrates the cell-grid after the first drop is deposited at cell (1,1), with solid lines representing the region of interest for the first drop, $B_1$, and blue disk representing the drop;
  (c) illustrates the cell-grid after the second drop is deposited, with grey disk representing the existing drop and same region and new drop as in (a);
  (d) illustrates the cell-grid after the third drop is deposited.}
  \label{fig:schematics}
\end{figure*}

\section{\MakeUppercase{Modeling height profile of multiple drops in a straight line}}
\label{sec:model}
The proposed model assumes that there is no coalescence between adjacent drop and the substrate is non-porous and flat.
No coalescence is guaranteed by exposing the drop under UV light right after each deposition.
The substrate is a microscope glass slide,
which satisfies aforementioned requirements.
For a sequentially printed straight line,
only the first drop is deposited on the flat substrate.
Since drops overlap, all subsequent drops are partially deposited on the prior drop,
which makes the effective substrate profile not flat. 
Thus, each subsequent drop flows from higher to lower locations.
This process involving flow speed, direction, distance and amount of ink. 
Known parameters include surface tension, viscosity, substrate height difference.
The model of this process will inevitably be very complicated.
Since the most common scenario is printing with same ink on the same substrate, 
which means surface tension and viscosity do not change throughout the process,
the model is simplified with the assumption that the ink flow only depends on the height difference along the printing path and it does not flow outside of the $3 \times 3$ cells.
The propagation of height profile is indirectly calculated with the volume and area, i.e.
\begin{equation}
    \centering
    \label{eq:hnew}
    h_{i j}[k] = c \frac{v_{i j}[k-1]+\Delta v_{i j}[k]}{a_{i j}[k-1]+\Delta a_{i j}[k]},
\end{equation}
where $v_{i j}[k-1]$ and $a_{i j}[k-1]$ are the prior volume and area within cell $(i,j)$, respectively;
$\Delta v_{i j}[k]$ and $\Delta a_{i j}[k]$ are the change of volume and change of area due to the $k^{th}$ drop at cell $(i,j)$, respectively;
and $c$ is the ratio between single drop volume $(\mu m^3)$ and single cell area $(\mu m^2)$. 

Figure \ref{fig:schematics} illustrates the cells,
region of interests,
existing drop and new drop. 
It is based on printing a 4-drop straight line. 
To characterize 4 drops, a $3\times 5$ cell-grid, 
shown in \mbox{Fig. \ref{fig:schematics}.(a)}, is needed.
The $k^{th}$ drop is deposited at the center of cell (1,$k)$.
Since a drop is assumed to affect $3 \times 3$ cells,
the region of interest of the $k^{th}$ drop is denoted as $B_k$, 
which is a set of cells, $\{(i,j)|i\in \{0,1,2\}, j\in \{ k-1,k,k+1\}\}$.
For the first drop, the region of interest and the drop is shown in \mbox{Fig. \ref{fig:schematics}.(b)},
where the blue disk represents the drop and black solid grids represent the region of interest.
After the first drop is deposited and cured,
part of it will be the new substrate for the second drop.
This is shown in \mbox{Fig. \ref{fig:schematics}.(c)}, where the gray disk represents the cured first drop.
The figure illustrates the moving region of interest.
Similarly, Fig.\ref{fig:schematics}.(d) shows the cured drops, the region of interest and the third drop.

Mathematically, the volume,
area and height matrices of the $k^{th}$ drop depositing at cell $(1,k)$ can be written as 
\begin{equation}
    \label{eq:vk}
    V[k] = \left [v_{ij}[k]|(i,j)\in B_k \right]= \left [
    \begin{array}{ccc}
      v_{0k-1}[k]  & v_{0k}[k] & v_{0k+1}[k] \\
      v_{1k-1}[k]  & v_{1k}[k] & v_{1k+1}[k] \\
      v_{2k-1}[k]  & v_{2k}[k] & v_{2k+1}[k]
    \end{array} \right],
\end{equation}
\begin{equation}
    \label{eq:ak}
    A[k] = [a_{ij}[k]|(i,j)\in B_k ]= \left [
    \begin{array}{ccc}
      a_{0k-1}[k]  & a_{0k}[k] & a_{0k+1}[k] \\
      a_{1k-1}[k]  & a_{1k}[k] & a_{1k+1}[k] \\
      a_{2k-1}[k]  & a_{2k}[k] & a_{2k+1}[k]
    \end{array} \right],
\end{equation}
and 
\begin{equation}
    \label{eq:hk}
    H[k] = [h_{ij}[k]|(i,j)\in B_k ]= \left [c \left. \frac{v_{ij}[k]}{a_{ij}[k]} \right|(i,j)\in B_k \right].
\end{equation}

\subsection{\MakeUppercase{Volume Propagation}}
\label{sec:vol_model}

With $\Delta V[k]$ denoting the change of volume due to the $k^{th}$ drop within region of interest,
the associated volume propagation can be written as
\begin{equation}
    \label{eq:vol_propagation}
    \begin{split}
        V[k] &= V[k-1]S + \Delta V[k] \\
        \Delta V[k] &= f(H[k-1], M_v[k], u[k]) \\
         & = H[k-1](S-I)M_v[k] u[k],
    \end{split}
\end{equation}
where $S$ is a shift matrix that shifts the second and the third column of a matrix one column to the left and fills the now empty third column with 0, 
it can be written as
\begin{equation}
    \label{eq:s_matrix}
    S = \left[ 
    \begin{array}{ccc}
        0 & 0 & 0 \\
        1 & 0 & 0 \\
        0 & 1 & 0
    \end{array}\right];
\end{equation}
$M_v[k]$ is a $3 \times 3$ matrix characterizing the impact of height difference on volume distribution of the $k^{th}$ drop;
and $u[k]$ is the binary  input. 
$u[k] =1 $ if a drop is deposited and $u[k]=0$ if no drop is deposited. 
$M_v[k]$ can be calculated from 
\begin{equation}
    \label{eq:mv}
    M_v[k] = (S-I)^{-1}H[k-1]^{-1}\Delta V[k],
\end{equation}
when $u[k]$ is 1.



\subsection{\MakeUppercase{Area Propagation}}
\label{sec:area_model}
The area propagation is derived in a similar fashion as the volume propagation.
The only difference is that the cell area cannot be greater than 1,
which means 100\% occupied.
As a result,
with $\Delta A[k]$ denoting the change of area due to the $k^{th}$ drop within region of interest,
the area propagation law can be written as
\begin{equation}
    \label{eq:area_propagation}
    \begin{split}
        A[k] &= A[k-1]S + \Delta A[k] \\
        \Delta A[k] &= \text{min} \{g(H[k-1], M_a[k], u[k]),1\} \\
         & = \text{min} \{H[k-1](S-I)M_a[k] u[k],1\},
    \end{split}
\end{equation}
where $M_a[k]$ is a $3 \times 3$ matrix characterizing the impact of height difference on area, similar to $M_v[k]$.
$M_a[k]$ can be calculated from 
\begin{equation}
    \label{eq:ma}
    M_a[k] = (S-I)^{-1}H[k-1]^{-1}\Delta A[k]
\end{equation}
when $u[k]$ is 1. 


\section{\MakeUppercase{Experimental Validation}}
\label{sec:exp_valid}
Experiments are conducted to obtain coefficients and validate the model.
$M_v[k]$ and $M_a[k]$ are obtained from one group of experimental data with \mbox{Eq. (\ref{eq:mv})} and \mbox{Eq. (\ref{eq:ma})}, respectively.
The experimentally obtained $M_v[k]$ and $M_a[k]$ are used in the model.
Subsequent model validation is done by comparing the root mean square (RMS) error of the height profile.
The RMS height profile error is given by
\begin{equation}
    \centering
    \label{eq:rms}
    RMS = \sqrt{\frac{1}{9} \sum_{i=0}^2 \sum _{j=k-1}^{k+1} \left (\frac{\bar{h}_{ij}[k]-h_{ij}[k]}{\bar{h}_{ij}[k]} \right)^2},
\end{equation}
where $\bar{h}_{ij}[k]$ represents the average height at cell $(i,j)$ of all samples 
and $h_{ij}[k]$ represents the predicted height at cell $(i,j)$ after the $k^{th}$ drop is deposited.

Printed lines of 1 to 6 drops in length at 90 $\mu m$ pitch distance are done with 75 duplicates.
Height and drop images ae collected for all the printed lines. 

\subsection{\MakeUppercase{Obtain} $M_v[k]$ AND $M_a[k]$}
\label{sec:cal_mv_ma}

To avoid coalescence, a drop is not deposited if there is uncured ink within region of interest.
Each drop is cured under UV light for 1 second. 
Samples are measured after the experiment session.
With 75 samples for each pattern, we randomly separate samples of each pattern to 3 groups of 25 samples.

\begin{figure}
    \centering
    \includegraphics[width=\linewidth]{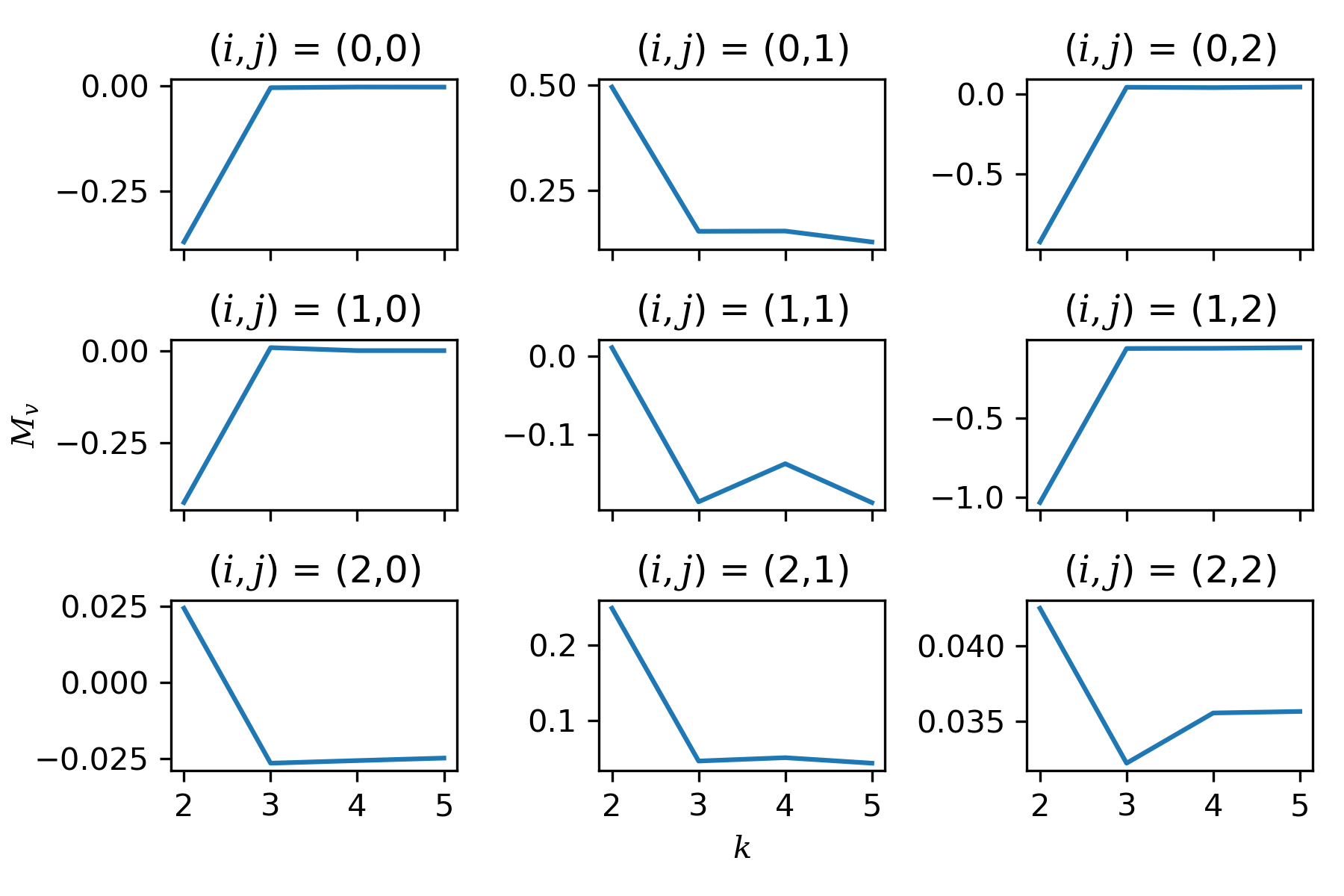}
    \caption{$M_v[k]$ calculated from \mbox{Eq. (\ref{eq:mv})} are plotted as a function of $k$.
    It can be observed that $M_v$ is approximately constant starting from the third drop.}
    \label{fig:mv_trend}
\end{figure}

\begin{figure}
    \centering
    \includegraphics[width=\linewidth]{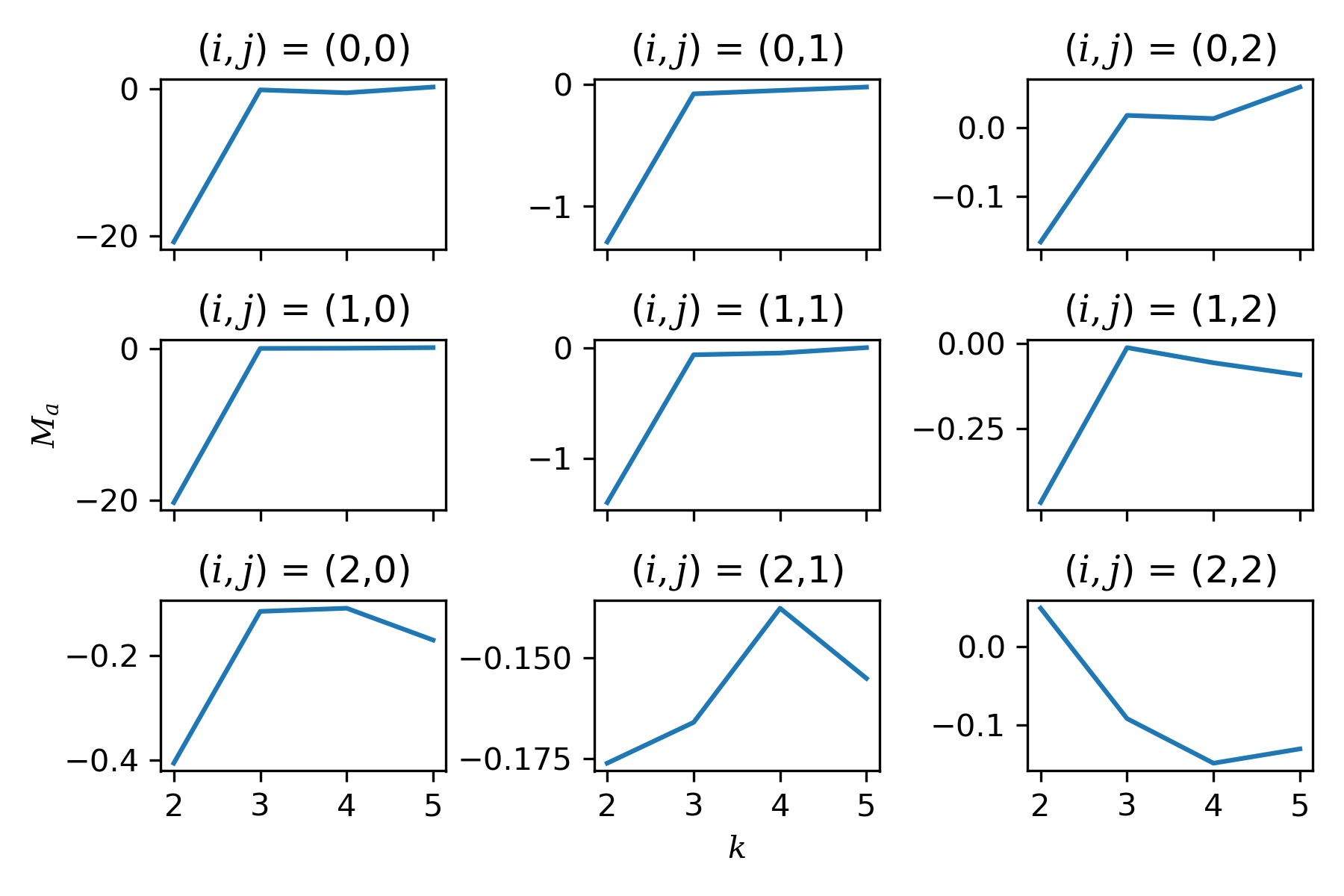}
    \caption{$M_a[k]$ calculated from \mbox{Eq. (\ref{eq:ma})} are plotted as a function of $k$.
    It can be observed that $M_a$ is approximately constant starting from the third drop.
    All $M_a[k]$ are in a small range with $(i,j)=(2,1)$.}
    \label{fig:ma_trend}
\end{figure}

Section \ref{sec:vol_model} and Section \ref{sec:area_model} introduced the model of both volume and area propagation.
To determine $M_v[k]$ and $M_a[k]$, 
one group of samples are used to calculate both $M_v[k]$ and $M_a[k]$.
Both $M_v[k]$ and $M_a[k]$ are plotted as a function of $k$, 
shown in \mbox{Fig. \ref{fig:mv_trend} and \ref{fig:ma_trend}}, respectively.
As shown in \mbox{Fig. \ref{fig:mv_trend} and \ref{fig:ma_trend}}, 
both $M_v[k]$ and $M_a[k]$ show a similar trend.
$M_v[k]$ and $M_a[k]$ can be approximated as a constant  for $k\ge 3$.
From experimental data, 
$M_v$ and $M_a$ are calculated from \mbox{Eq. (\ref{eq:mv})} and \mbox{Eq. (\ref{eq:ma})}, respectively.
The results are shown in 
\begin{equation}
\centering
\label{eq:mv_value}
M_v = \left[
\begin{array}{rrr}
    0.0057 & 0.1526 & 0.043 \\
    0.0097 & -0.1853 & -0.0569 \\
    -0.0266 & 0.0271 & 0.0032
\end{array} \right],
\end{equation}
and
\begin{equation}
\centering
\label{eq:ma_value}
M_a = \left[
\begin{array}{rrr}
    -0.1202 & -0.0797 &  0.0179 \\
    0.0574 & -0.0612 & -0.1294 \\
    -0.1150 & -0.1660 & -0.0921
\end{array} \right].
\end{equation}

\begin{figure}
	\centering
	\includegraphics[width=\linewidth]{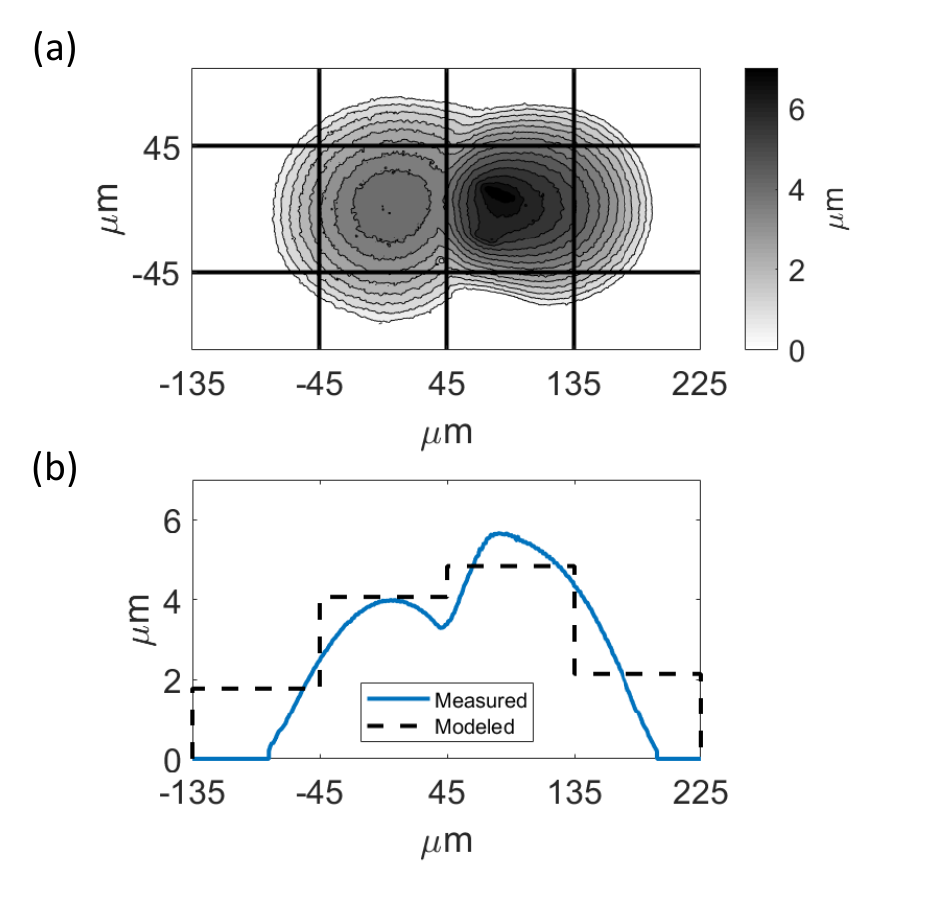}
	\caption{(a): Contour of 2 drops, marked with 90 $\mu m$ pitch size. \hspace{\textwidth}
	(b): The blue solid line represents the mean height of each column along the center row of the sample in (a); 
	black dashed line represents the cell height long center row from the model.}
	\label{fig:2_drop}
\end{figure}

\begin{figure}
	\centering
	\includegraphics[width=\linewidth]{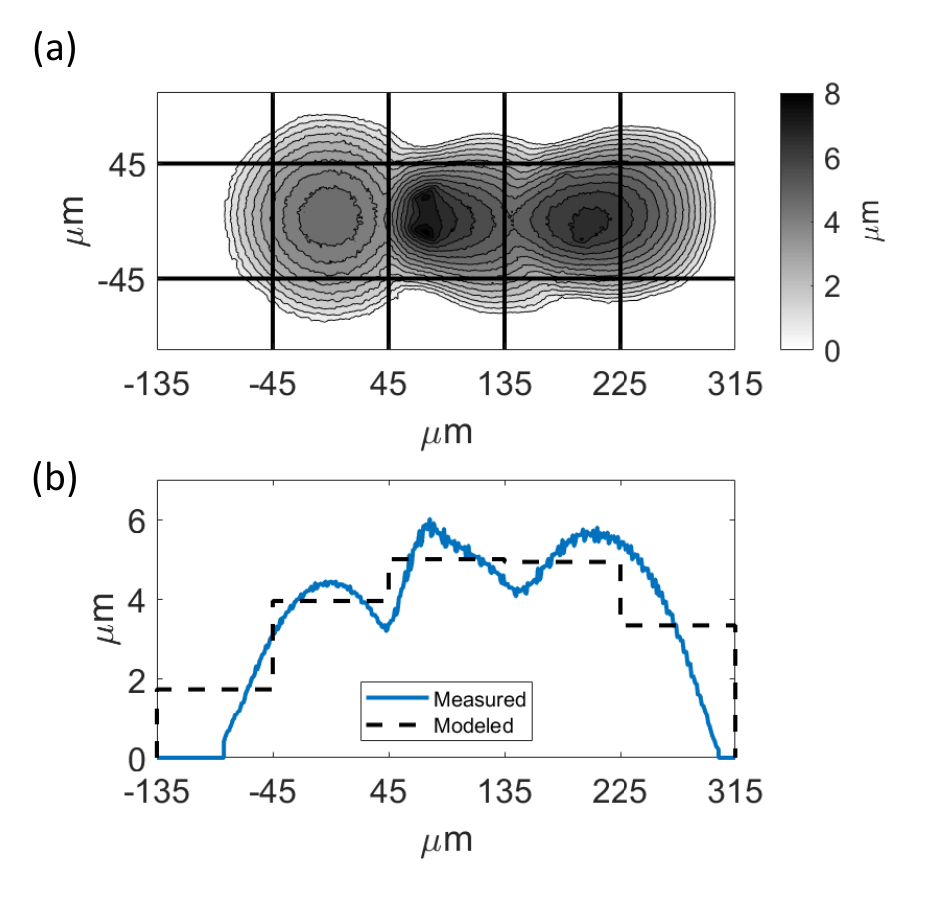}
	\caption{(a): Contour of 3 drops, marked with 90 $\mu m$ pitch size. \hspace{\textwidth}
	(b): The blue solid line represents the mean height of each column along the center row of the sample in (a); 
	black dashed line represents the cell height long center row from the model.}
	\label{fig:3_drop}
\end{figure}

Since one group of samples are used to show the trends in \mbox{Fig. \ref{fig:mv_trend} and \ref{fig:ma_trend}},
the other two groups are used for the modeling and validation. 
2-drop and 3-drop measurements are shown in Tables \ref{tab:prof_2} and \ref{tab:prof_3}, respectively.
Measurements are the average of all 25 samples in one group.
The shaded region in each table shows the footprint of one samples selected from each group
to help visualize the area covered by ink in each cell qualitatively.
Figures \ref{fig:2_drop}.(a) and \ref{fig:3_drop}.(a) show the contour of the 2-drop and 3-drop sample. 
Figures \ref{fig:2_drop}.(b) and \ref{fig:3_drop}.(b) show the comparison between the mean measured height and the model computed cell height along the center row of the corresponding sample in (a).
Consistent with cells in \mbox{Fig. \ref{fig:2_drop} and \ref{fig:3_drop}}, 
solid lines represent the cell boundaries. 

\begin{table}[]
	\centering
	\caption{Measured volume, area and height $(\mu m)$ in each cell of 2 drops.}
	\label{tab:prof_2}
	\includegraphics[height=12em]{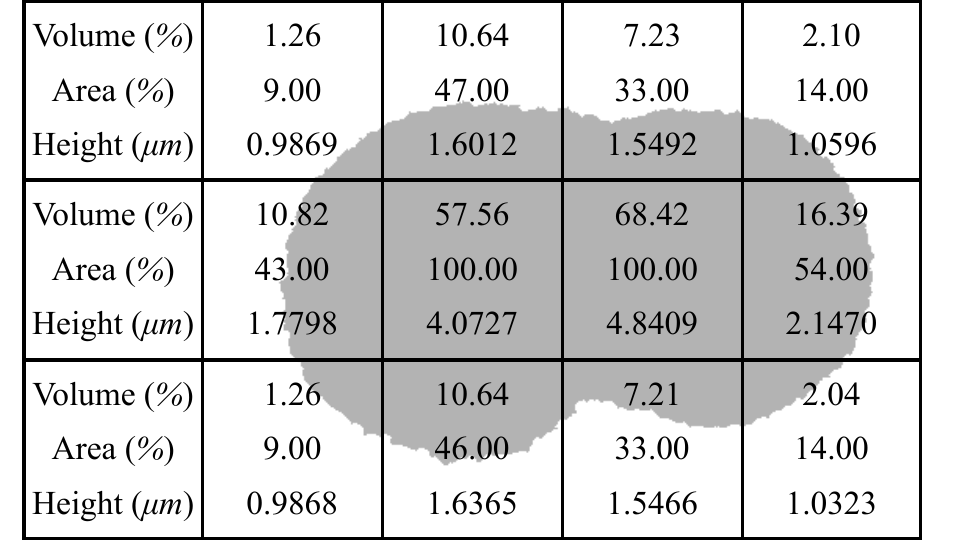}
\end{table}

\begin{table}[]
	\centering
	\caption{Measured volume, area and height $(\mu m)$ in each cell of 3 drops.}
	\label{tab:prof_3}
	\includegraphics[height=12em]{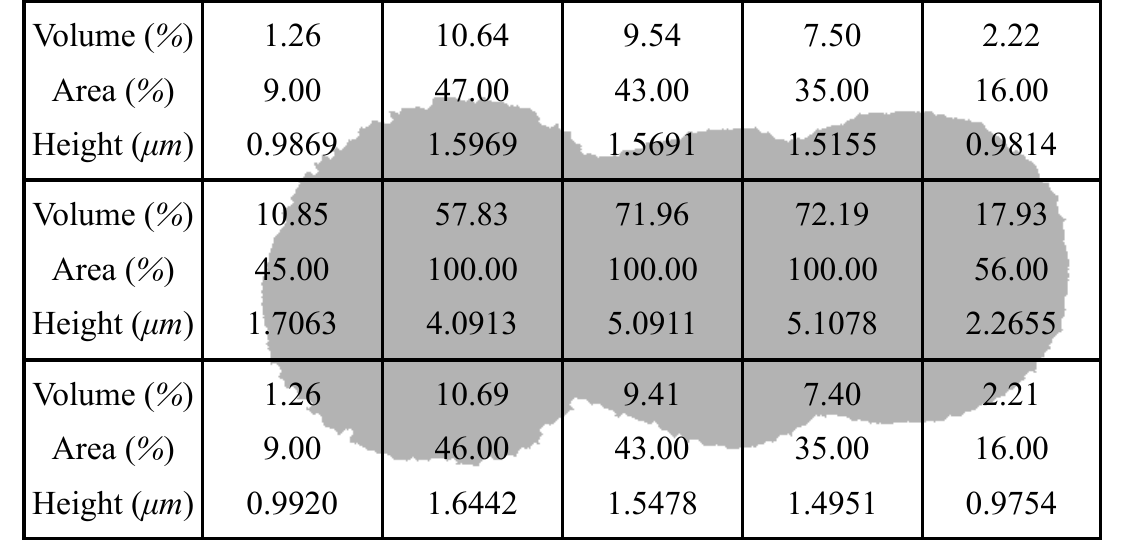}
\end{table}

\subsection{\MakeUppercase{Results of 4-drop and more Pattern}}
\label{sec:result}
With $M_v$ and $M_a$ determined,
lines with 4 and more drops are used to verify the model and analyze the errors.
Predictions of volume, area and height are obtained from \mbox{Eq. (\ref{eq:vol_propagation})}, \mbox{Eq. (\ref{eq:area_propagation})} and \mbox{Eq. (\ref{eq:hk})}, respectively.
Benchmark is RMS error,
which can be calculated from \mbox{Eq. (\ref{eq:rms})}.

\begin{table}[]
	\centering
	\caption{RMS errors of height prediction.}
	\label{tab:rms_prediction}
	\begin{tabular}{|c|c|c|c|c|}
		\hline
        \# of drops & 4 & 5 & 6 & 7 \\
        \hline
        Height Error & 5.9\% & 7.2\% & 6.4\% & 6.9\% \\
        \hline
	\end{tabular}
\end{table}

\begin{figure}
	\centering
	\includegraphics[width=\linewidth]{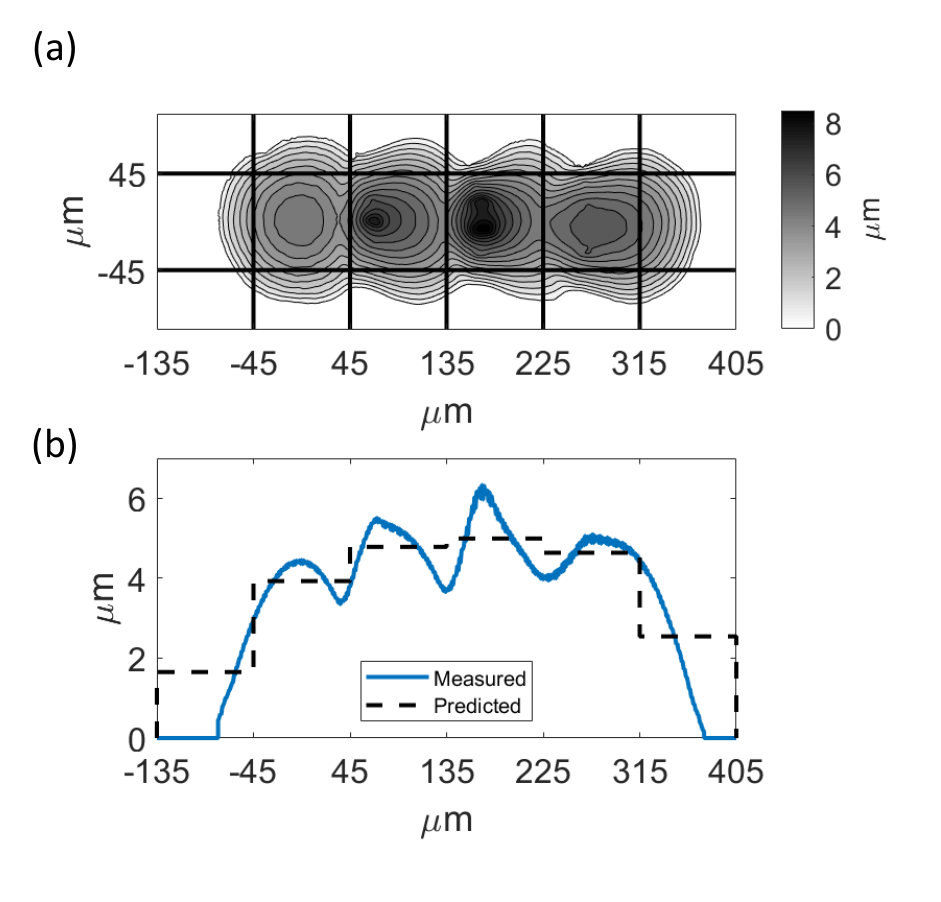}
	\caption{(a): Contour of 4 drops, marked with 90 $\mu m$ pitch size. \hspace{\textwidth}
	(b): The blue solid line represents the mean height of each column along the center row of the sample in (a); 
	black dashed line represents the cell height long center row predicted from the model.}
	\label{fig:4_drop}
\end{figure}

\begin{figure}
	\centering
	\includegraphics[width=\linewidth]{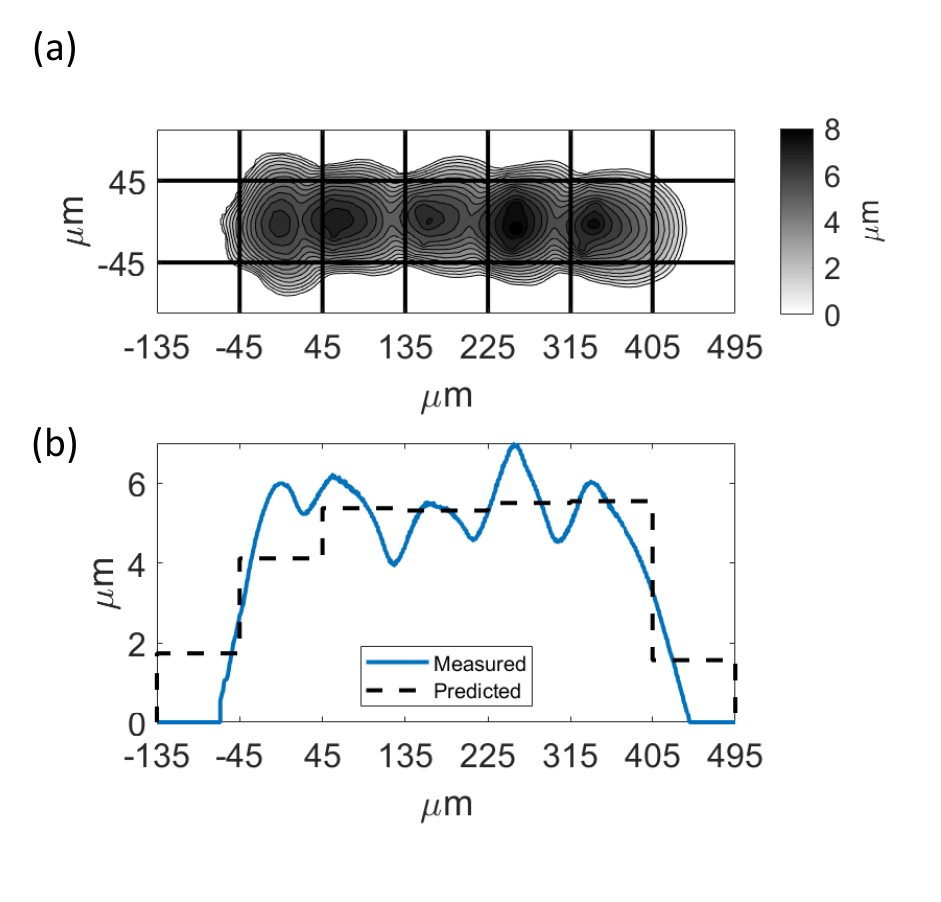}
	\caption{(a): Contour of 5 drops, marked with 90 $\mu m$ pitch size. \hspace{\textwidth}
	(b): The blue solid line represents the mean height of each column along the center row of the sample in (a); 
	black dashed line represents the cell height long center row predicted from the model.}
	\label{fig:5_drop}
\end{figure}

Figure \ref{fig:4_drop} shows the comparison between one sample of a 4-drop line and predicted height along the center row obtained from the model.
Figure  \ref{fig:4_drop}.(a) shows the contour of the sample with marked cell boundaries. 
The trend of ink spreading towards right is noticeable from last two cells from the right.
Figure  \ref{fig:4_drop}.(b) shows the comparison among the measured and predicted cell height along the center row.
The measured data are the mean of each column between -45 and 45 $\mu m$ in the corresponding (a) figure.
Figure \ref{fig:5_drop} is similar to \mbox{Fig. \ref{fig:4_drop} }with 5 drops. 
RMS errors of the height prediction for each pattern is shown in \mbox{Table \ref{tab:rms_prediction}}. 
All errors are less than 10 \%.

Table \ref{tab:comp} compares this model with the previous model and graph-based model on the prediction of 4-drop straight line.
Based on the same benchmark of RMS errors, 
this model has the lowest error while the graph-based model has the highest error of over 15\%.

\begin{table}[]
    \centering
    \caption{RMS error comparison of 4-drop line among 3 models}
    \label{tab:comp}
    \begin{tabular}{|c|c|c|c|}
        \hline
         & Graph-based & Previous & This Model \\ \hline
         RMS Error & 17.2\% & 7.4\% & 5.9\% \\ \hline
    \end{tabular}
\end{table}

\section{\MakeUppercase{Conclusion}}
\label{sec:conclusion}
In this paper, an improved model for drop-on-demand printing of UV curable inks is proposed. 
The model is based on the localized ink flow caused by the height difference of the substrate.
Height profile is indirectly obtained from volume and area propagation.
The propagation law of volume uses $M_v$ to quantify the impact of such height difference on volume distribution.
Similarly, the propagation law of area uses $M_a$ to quantify the impact of height difference on spreading ink.

Extended from our previous effort, which uses constant volume and area profile to calculate a height profile,
the proposed model uses second and third drop profile to determine $M_v$ and $M_a$ with \mbox{Eq. (\ref{eq:mv})} and \mbox{Eq. (\ref{eq:ma})},
respectively.
Height profiles of 4 and more drops are predicted from the model, 
and compared with experimental results. 
The proposed model is able to reduce the RMS height profile error more than 60\% from existing graph-based model and further improve from our previous model \cite{WuY.2019Mhpf}.
\bibliographystyle{plain}
\bibliography{library.bib}
\end{document}